\documentclass[11pt,letterpaper]{article}

\usepackage{graphicx}
\usepackage{fullpage}
\usepackage{multirow}
\usepackage{array}
\usepackage{amsmath}
\usepackage{dsfont}
\usepackage{wrapfig} 
\usepackage{sectsty}
\usepackage{setspace}
\usepackage[colorlinks=true,
            urlcolor=blue,
            citecolor=red,
            pdfstartview=FitH,
            pdfpagemode=None]{hyperref}

\usepackage[square,numbers,comma,sort&compress]{natbib}

\sectionfont{\small}

\newcommand\CNOT{\ensuremath{\textit{CNOT\/}}}
\newcommand\CPHASE{\ensuremath{\textit{CPHASE\/}}}
\renewcommand{\>}{\rangle}

\input{Qcircuit}

\title{The impact of classical electronics constraints on a solid-state
logical qubit memory}

\author{
James E. Levy\thanks{Sandia National Laboratories \{\texttt{jelevy}, \texttt{aganti}, \texttt{caphill},
\texttt{brhamle}, \texttt{tmgurri}, \texttt{rdcarr}, \texttt{mscarro}\}
\texttt{@sandia.gov}.  Sandia is a multipurpose laboratory operated by
Sandia Corporation, a Lockheed-Martin Company, for the United States
Department of Energy under contract DE-AC04-94AL85000.}
\and
Anand Ganti\footnotemark[1]
\and
Cynthia A. Phillips\footnotemark[1]
\and
Benjamin R. Hamlet\footnotemark[1]
\and
Andrew J. Landahl\thanks{Center for Advanced Studies, Department of Physics and Astronomy, University of New Mexico, Albuquerque, NM, USA, {\tt alandahl@unm.edu}.}
\and
Thomas M. Gurrieri\footnotemark[1]
\and
Robert D. Carr\footnotemark[1]
\and
Malcolm S. Carroll\footnotemark[1]
}

\begin{document}

\maketitle

\newcommand {\cnot}{\emph{CNOT}\ }
\newcommand {\cphase}{\emph{CPHASE}\ }
\newcommand {\swap}{\emph{SWAP}\ }
\newcommand {\mxor}{\emph{XOR}\ }
\newcommand {\mnot}{\emph{NOT}\ }

\newcommand{\todo}[1]{\vspace{5 mm}\par \noindent \marginpar{\textsc{ToDo}}
\framebox{\begin{minipage}[c]{0.95 \columnwidth} \tt #1
\end{minipage}}\vspace{5 mm}\par}


\begin{abstract}
We describe a fault-tolerant memory for an error-corrected
logical qubit based on silicon double quantum dot physical qubits.
Our design accounts for constraints imposed by supporting classical
electronics.  A significant consequence of the constraints is to add
error-prone idle steps for the physical qubits.  Even using a schedule
with provably minimum idle time, for our noise model and choice of
error-correction code, we find that these additional idles negate any
benefits of error correction.  Using additional qubit operations, we
can greatly suppress idle-induced errors, making error correction
beneficial, provided the qubit operations achieve an error rate less
than $2 \times 10^{-5}$.  We discuss other consequences of these
constraints such as error-correction code choice and physical qubit
operation speed.  While our analysis is specific to this memory architecture,
the methods we develop are general enough to apply to other
architectures as well.

\end{abstract}


\section{Introduction}
\label{sec:intro}
Quantum information processing (QIP) promises a path towards resolving
currently computationally-intractable problems \cite{Nielsen:2000a}.
However, quantum bits (qubits) used for storing quantum information
are, unfortunately, much more susceptible to errors than classical
bits.  Realization of error-corrected quantum computation, therefore,
represents a critical QIP engineering pursuit.  A key concept in this
pursuit is the redundant encoding of a {\em logical qubit} in the state of
many physical qubits.  This redundancy allows one to check for errors
and correct them.

This paper presents a solid-state architecture for a single logical
qubit memory that accounts for the constraints imposed by both
\emph{\textbf{classical electronics}} and the \emph{\textbf{native
quantum gate set}}---the available set of qubit transformations the
solid-state system provides.  Quantum-computing architectures have
been considered previously, for example in ion traps~\cite{Balensiefer:2005a} and
solid-state~\cite{Loss:1998a, Hollenberg:2006a}.  These analyses began the study of
incorporating realistic implementation constraints. We extend these
studies in the solid-state to include explicit electronic constraints,
where we expect electronics integration to be easier and the least
constrained.  From this we have gained a number of critical
architectural insights including guidance about error-correcting code
choice and quantum gate speed and scheduling limitations.

The rest of this paper is structured as follows. 
Section~\ref{sec:QCBackground} is a brief background on quantum computing. 
Section~\ref{sec:physicalQubit} describes the solid-state qubit implementation, 
its assumed constraints and assumptions about noise.  Section~\ref{sec:localCheckCodes}
provides an introductory description of quantum error 
correction and specifics on the Bacon-Shor code. Section~\ref{sec:logicalQubit} 
describes an ``enclosed" architecture for a fault-tolerant logical qubit that is 
matched to the solid-state qubit constraints,  and describes classical 
electronics constraints for scheduling error correction operations within the 
logical qubit. Section~\ref{sec:IP} describes how we optimized the error-correction
schedule subject to these constraints. Section~\ref{sec:threshold} 
quantifies when it is beneficial to use this logical qubit with a given native 
gate set and schedule.  Section~\ref{sec:conclusion} summarizes our results and concludes.


\section{Quantum computing background}
\label{sec:QCBackground}
Every quantum computation can be expressed as a \emph{quantum circuit} in
which a sequence of elementary transformations called \emph{gates} act on a
collection of elementary parcels of information called \emph{qubits}.  To
understand the basics of quantum computing, then, it suffices to understand
what qubits and gates are and how they interact.  For a more detailed
treatment of quantum computation, we refer the reader to any of the growing
number of textbooks on the subject, such as Ref.~\cite{Nielsen:2000a}.

A qubit is formed out of two isolated quantum states (\textit{e.g.}, a
ground and excited state of an atom).  Mathematically, we represent a
qubit's two states by the vectors $|0\> := \binom{1}{0}$ and $|1\> :=
\binom{0}{1}$, which we call \emph{computational basis states}.  A general
single-qubit state is of the form $\cos \frac{\theta}{2}|0\> +
e^{i\varphi}\sin \frac{\theta}{2}|1\>$, where $\theta$ and $\varphi$ are
polar and azimuthal angles in spherical coordinates; the set of possible
single-qubit states forms what is called the \emph{Bloch sphere}.

A single-qubit (coherent) gate can be expressed as a rotation of the Bloch
sphere.  Some special one-qubit gates that we will discuss are the bit-flip
gate $X$, phase-flip gate $Z$, and the Hadamard gate $H$, which correspond
to rotations by $\pi$ about the axes $\hat{x}$, $\hat{z}$ and
$\frac{1}{\sqrt{2}}(\hat{x} + \hat{z})$ respectively.  Instead of being
thought of in terms of abstract rotations, these gates are probably best
understood in terms of their action on computational basis states:
\begin{align}
X|0\> &= |1\> &
Z|0\> &= |0\> &
H|0\> &= |+\> := \frac{1}{\sqrt{2}}(|0\> + |1\>) \\
X|1\> &= |0\> &
Z|1\> &= -|1\> &
H|1\> &= |-\> := \frac{1}{\sqrt{2}}(|0\> - |1\>).
\end{align}

Mostly we will be interested in the rotation gates $X_\theta$ and $Z_\theta$
about the $\hat{x}$ and $\hat{z}$ directions by arbitrary angles $\theta$;
conveniently the Hadamard gate can be expressed as the triple sequence of
$\pi/2$ rotations about $\hat{z}$, then $\hat{x}$, then $\hat{z}$, or more
succinctly, $H = Z_{\pi/2}X_{\pi/2}Z_{\pi/2}$.

A measurement of a qubit in the computational basis, an operation we denote
by $M_Z$, will transform an isolated qubit's state to $|0\>$ with
probability $\cos^2 \frac{\theta}{2}$ or to $|1\>$ with probability $\sin^2
\frac{\theta}{2}$.  More generally, a qubit can be correlated with other
qubits and the outcome probabilities of measurements will reflect these
correlations.  Because $M_Z$ transforms a qubit, we will sometimes also call
this a one-qubit (incoherent) gate.  In principle one can measure a qubit in
any basis; such a measurement is equivalent to a rotation which takes the
computational basis to that basis, followed by $M_Z$.  For example, a
measurement consisting of $|+\>$ and $|-\>$ is denoted $M_X$, and can be
thought of as performing a Hadamard gate followed by $M_Z$.

Given a one-qubit (coherent) gate $U$, the two-qubit \emph{controlled-$U$
gate}, denoted $\Lambda(U)$, is defined by
\begin{align}
\Lambda(U)|c\>|\psi\> :=
%
%
\left\{ \begin{matrix}
      |0\>|\psi\>  & \text{if } c=0 \\
      |1\>\,U|\psi\>  & \text{if } c=1.
\end{matrix} \right.
\end{align}

The only two-qubit controlled gates we will discuss are the the
\emph{controlled-NOT} gate $\CNOT := \Lambda(X)$ and the
\emph{controlled-phase} gate $\CPHASE := \Lambda(Z)$.
While many other two-qubit gates exist, the only other ones we consider are 
\emph{Pauli operators}.  The set of one-qubit Pauli operators are $X$, $Z$, 
$XZ$, and the identity gate $I$.  The set of two-qubit Pauli operators are the 
two-qubit gates that act as a one-qubit Pauli operator on each qubit.  The 
reason these gates are interesting is that any noise process on one qubit can be 
expressed as a linear combination of one-qubit Pauli operators and every noise 
process on two qubits can be expressed as a linear combination of two- qubit 
Pauli operators.  In Sec. 5, we will discuss a noise model in which gates are 
assumed to act ideally followed by a Pauli operator drawn at random.

A universal gate set is necessary to do arbitrary computations.  It is
well-known that there are certain sets of classical logic gates over which
any Boolean function can be expressed.  An example of such a \emph{universal
gate basis} is the set consisting of \textsc{NAND} and \textsc{FANOUT}.  In
an analogous way, there are quantum universal gate bases over which any
multi-qubit transform $U$ can be efficiently approximated.  An example is
$\{ H, Z_{\pi/4}, Z_{\pi/2}, \CNOT, M_Z \}$, sometimes called the
\emph{standard gate basis}.  To perform quantum error-correction, though,
one does not need a fully-universal gate basis.  It suffices to use only
\emph{Clifford circuits}, namely those generated by the gates $H$,
$Z_{\pi/2}$, $\CNOT$, and $M_Z$.  Clifford circuits are also generated by
the set $\{ |0\>, |1\>, M_Z, M_X, \CNOT \}$, where $|0\>$ and $|1\>$ in this
context represent deterministic gates that prepare the states $|0\>$ and
$|1\>$ respectively.


\section{Physical Qubit, Native Gate Set, and Noise Assumptions for Logical Qubit}
\label{sec:physicalQubit}
The physical qubit for this logical qubit analysis is a two electron spin system. 
The two electron spins are confined within a silicon double quantum dot (Si DQD) 
and they form two distinct spin configurations, a singlet or triplet, 
with two distinct energies analogous to a ground and excited state. 
These two states form the qubit's computational basis states $|0\>$ and $|1\>$. 
A \emph{gate set} for the DQDs in GaAs was proposed by Taylor \textit{et al.} \cite{Taylor:2005a},
which consists of $\{|1\>, M_Z, X_\theta, Z_\theta, \CPHASE\}$, Table~\ref{tab:gate-params}.  
For effecting quantum error-correction,
only a finite subset of these gates are necessary, for example the set
$\{|1\>, M_Z, Z_{\pi/2}, Z, X_{\pi/2}, X, \CPHASE \}$ suffices. We define this set of
gates as the Si DQD \emph{native gate set}.  In addition to these gates,
qubits also experience the ``identity gate'' $I^*$ by sitting idle.  The $I$ gate is
a modified idle, explained below, that relies on additional gate pulses to suppress noise.   

	\begin{table}
	\centering
	\caption{Gate Times and Failure Probabilities. $\tau=30$ ns. $p=0.3\%$.}
  \label{tab:gate-params}
		\begin{tabular}{| c | c | p{.75in} | p{.8in} | p{.7in} | p{2.0in} |}
		\hline
		Gate & Time & Failure Probability & Estimated Decay Time & Estimated Error & Dominant Noise Source \mbox{(anticipated)}\\
		\hline \hline
		$M_Z$, prep\,$|1\rangle$  & $\tau$   & $p/30$   & N/A            & $1\times10^{-4}$     & pulse error $\Rightarrow$ incomplete or non-adiabatic transition \\  \hline
		$S=Z_{\pi/2}$             & $\tau$   & $p$      & $\sim10\,\mu$s  & $3\times10^{-3}$     & charge fluctuations on $J$ \\ \hline
		$Z=Z_{\pi}$               & $2\tau$  & $2p$     & $\sim10\,\mu$s  & $6\times10^{-3}$     & charge fluctuations on $J$ \\ \hline
		$X_{\pi/2}$               & $3\tau$  & $4p$     & $\sim10\,\mu$s  & $1.2\times10^{-2}$     & charge fluctuations on $J$ \\ \hline
		$X=X_{\pi}$               & $4\tau$  & $4p$     & $\sim10\,\mu$s  & $1.2\times10^{-2}$   & charge fluctuations on $J$ \\ \hline
		\cphase                   & $4\tau$  & $4p$     & $\sim10\,\mu$s  & $1.2\times10^{-2}$   & charge fluctuations on $J$ \\ \hline
		$I^*$                     & $\tau$   & $9.99\times10^{-3}$      &
$\sim3\,\mu$s   & $1\times10^{-2}$   & non-uniform $B$-field no DD \\ \hline
		$I$                       & $\tau$   & $5\times10^{-7}$ & $60\,$ms         & $5\times10^{-7}$    & time varying non-uniform $B$-field uncompensated by DD \\
		\hline
		\end{tabular}
	\end{table}
	
\begin{figure}[h]
\centering
\includegraphics[scale=0.5]{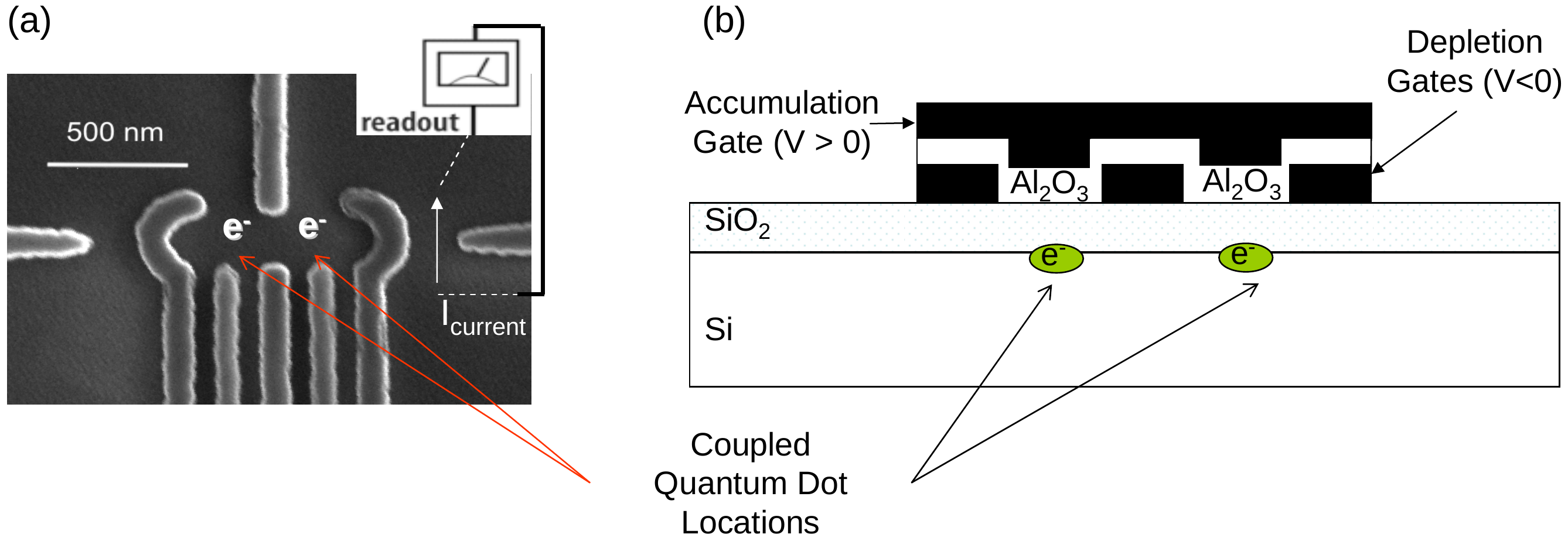}
\caption{Solid-state DQD. $(a)$ Top view SEM (Interconnects not all shown). $(b)$DQD cross-section.}
\label{fig:physQubit}
\end{figure}

The Si DQD structure is assumed to look analogous to the GaAs qubit described by 
Taylor \textit{et al.}, using the same metal routes and area to electrostatically define 
the dots, Fig.~\ref{fig:physQubit}~$(a)$, with one exception being a top metal 
gate, Fig.~\ref{fig:physQubit}~$(b)$.  The reservoir of electrons, out of which 
the single electron spins are isolated with the depletion gates, are produced by 
the application of a positive bias on the metal of a standard metal-oxide-
semiconductor stack, Fig.~\ref{fig:physQubit}~$(b)$, which draws electrons into 
the critical area. The GaAs qubit does not require this metal gate because the 
electron reservoir can be built-in. Figure~\ref{fig:physQubit}~$(a)$ also shows 
the inclusion of conducting routes that form a charge sensor (left and right
most gates), to measure the qubit state.  A total of 17 conducting routes 
are needed per Si DQD qubit not all of which are shown in 
Fig.~\ref{fig:physQubit}~$(a)$.  Routes not indicated in the figure are 4 ohmic 
contacts, 2 local inductor leads, the top gate, and 2 $\CPHASE$ enables. 

For the Si DQD qubit, we further modify the Taylor architecture in six key ways. 
We (1) identify the singlet, not the triplet, as the $|1\rangle$ state; (2) 
model measurement, $M_Z$, as being much faster and more accurate, assuming a 
recently proposed integrated read-out approach \cite{Gurrieri:2008a}; (3) use 
measurement for rapid initialization of the $|1\rangle$ state; (4) insert extra 
current pulses into the \textit{X}-gate to negate effects of stray magnetic 
fields on neighbor qubits; (5) do not use physical qubit transport mechanisms 
despite being assumed available in the Taylor architecture; and (6) insert, in 
some instances as described later, dynamical decoupling pulses to achieve very low 
error memory, $I$, compared to the bare memory $I^*$. Several of the proposed 
modifications warrant further clarification.  In the case of state preparation, 
error correction algorithms require the ability to prepare a qubit in a known 
state like the singlet, $|1\rangle$. The approach proposed for this logical 
qubit is to use measurement for state preparation by rapidly collapsing the 
state into a known $|0\rangle$ or $|1\rangle$ state rather than relying on triplet 
relaxation, $|0\rangle$ to $|1\rangle$ state, which is relatively slow. Although 
the error correction algorithm assumes a $|1\rangle$ starting state, a measurement 
that yields a $|0\rangle$ can be treated as a known error which can be accounted for 
later by classical feed-forward correction.  Feed-forwarding is faster and less error-prone than
correcting immediately with an \textit{X}-gate.  The extra current pulses 
proposed for the \textit{X}-gate are used in conjunction with additional $Z$-gates 
to exchange the electron positions within the double quantum dot (DQD) qubit, 
this allows the application of an opposite polarity $B$-field half way through 
the \textit{X}-gate rotation.  In this way, all neighbor qubits see a net zero 
rotation due to stray fields (assuming an identical but opposite $B$-field 
pulse), while the target qubit sees the intended net \textit{X} rotation. The 
challenges related to transport and our reasoning not to use it are explained 
later in this section.
 
A key contribution of this study is to examine the effect of electronics
impact on the overall logical qubit performance.  The details of the
physical qubit and the \emph{native gate set} define many of the
requirements for the electronics.  The effect of the \emph{native gate set}
and physical qubit on the electronics can be broken into its effect on three
classic architectural design trade-off areas \emph{power}, \emph{time}, and
\emph{space}.  \emph{Power} constraints are very strict in the solid-state
implementation because the DQD qubits must be cooled to $\sim$100\,mK.
Cooling to 100\,mK requires specialized cryogenics (i.e., dilution
refrigerators) that have very limited cooling powers especially at the
100\,mK stage.  This limits the amount of electronics that can be run on the
100\,mK stage and forces most active (power consuming) electronics up to
higher temperature stages, 4\,K or 300\,K, in the dilution refrigerator.
Details regarding the choice of the staging of the electronics will be
covered in Sec.~\ref{sec:logicalQubit}.

\emph{Time} of gates is impacted by the electronics through limits on timing 
precision (jitter), bandwidth between cryostat stages, and possibly cooling 
power (i.e., faster electronics dissipate more power).  
A clock period of 30\,ns was used in this work.  This clock period was 
chosen as a conservative estimate of what could be sustained with limited lines 
between stages, see Sec.~\ref{sec:logicalQubit}, while assuring high timing 
precision for the gates (i.e., jitter).  Scheduling constraints due to limits on 
the number of parallel operations are also discussed in Sec.~\ref{sec:logicalQubit}.

The \emph{space} required for the DQD qubit can have a severe affect on two 
dimensional lay-out at the 100\,mK stage through the relationship of the qubit 
size and number of metal routes required for each qubit (i.e., 17 metal lines 
and $\sim1\,\mu\text{m}^2$).  One of the challenges related to space is 
that there is a limit to how many neighboring qubits can be placed in a row 
without overextending the number of possible metal lines available in that row 
for a given CMOS process, see Sec.~\ref{sec:logicalQubit}.  The logical qubit 
is, therefore, constrained to more of a quasi-1D lay-out, which was also 
previously noted by Szkopek \textit{et al.} \cite{Szkopek:2006a}.  An important and weak 
assumption in this logical qubit is that each qubit is similar, which implies no 
additional tuning circuitry for individualized local tuning (tuning the DQD 
itself) or pulsing (tuning the pulse generators) is needed.  We expect serious additional 
space and time penalties for tuning circuitry which is a topic for 
future work.

\emph{Transport} of the physical qubit location can significantly relax 
\emph{space} and \emph{time} constraints.  However, current proposals for 
transport (e.g., shuttling \cite{Fujiwara:2001a}, and tunneling 
\cite{Switkes:1999a}) are even more speculative than the Si DQD and require 
additional hardware discovery. \emph{Logical transport} of information can 
alternatively be done through qubit operations like teleportation 
\cite{Nielsen:2000a} and $\swap$ (i.e., nearest neighbor exchange of qubit 
information without changing electron location).  However, neither of these 
operations are provided in the native gate set, so realizing them would require 
additional nontrivial and error-prone gating.  For example, the logical $\swap$ 
operation requires the translation of 3 $\CNOT$s into the native gate set.  We 
calculate that the relative error probability for a single $\swap$ operation 
using the DQD native gate set is 22$p$ and it takes 16 steps to execute, which 
is to be compared to other gates in Table \ref{tab:gate-params}.  The penalty 
for transporting by $\swap$ with this native gate set is, therefore, very high 
and undesirable.  

Noise and error correction, as previously noted, are a dominant issue for
quantum computations and are highly dependent on the choice of physical
qubit and native gate set.  In many cases, the effect of noise sources on a
qubit can be characterized with an empirical time constant fit to an
exponential decay in time \cite{Nielsen:2000a}.  The time constant can
describe probability of errors such as spin flips, $X$-like error, or
dephasing (e.g., transversal decoherence).  Most information about noise
sources for electron spins in silicon comes from measurements of spin
ensembles that are confined by donors~\cite{Tyryshkin:2003a}. These
measurements indicate very low probabilities of spin flips and relatively
slow transversal decoherence, denoted $T_2$, of
$\sim100\,\mu$s--$60\,$ms~\cite{Tyryshkin:2003a}. These decoherence times
are considerably longer than those measured in GaAs, $T_2\sim1.2\,\mu$s
\cite{Petta:2005a}, which is one of the primary motivations for studying Si
qubits.  The longer decoherence time in Si has been assigned to the ability
to remove nuclear spin hyperfine coupling to the electron spin in isotope
purified $^{28}$Si. 

Additional gate pulses called dynamic decoupling pulses must be incorporated
to achieve the reported long decoherence times in Si
\cite{Khodjasteh:2005a}. These pulses cancel noise from static non-uniform
$B$-field gradients over the entire sample.  Without dynamical decoupling
(DD), the decoherence times are anticipated to be $\sim 1\,\mu$s in
silicon \cite{Lyon:2008a}, which we use to infer the error probability of
the $I^*$ gate in Table~\ref{tab:gate-params}.  In the case of the Si DQD
qubit, a dynamic decoupling gate sequence of
\textit{Z}-$Idle$-\textit{Z}-$Idle$ could be used similar to what was chosen
for the GaAs qubit work \cite{Petta:2005a}.  $Idle$ is defined here as an
arbitrary number, $N$, of memory gates of 30\,ns duration (i.e., $N \times
I$).  A lower probability of memory error is used, $I$ for the case that
dynamic decoupling is used and the decoherence time within the pulses is
much slower, Table~\ref{tab:gate-params}. For initial computational ease we
used an \textit{X}-$Idle$-\textit{X}-$Idle$ schedule in this logical qubit
analysis, which represents a worse case because $X$ errors are greater than
$Z$ errors.

Gate operations can expose qubits to further sources of noise. In particular the 
$Z_\theta$-gate uses an externally controllable exchange energy, $J$, to rotate 
the qubit. The exchange energy is sensitive to the proximity of the two 
electrons, which is manipulated by external voltages that drive the electrons 
together or apart. The exchange energy is believed to be sensitive to charge 
fluctuations, which can alter the intended proximity and result in random 
rotation, which can rapidly dephase the qubit \cite{Hu:2006a}. We model $T_2$ 
for the $Z$ gate to be $10\,\mu$s, based on an assumed charge dephasing time of 
$200$ ns \cite{Gorman:2005a} and a $J$ sensitivity on local electrical potential 
of $0.01$ \cite{Hu:2006a}.  Because the $J$ interaction is also used to effect
$\CPHASE$ gates and to implement our magnetic-field-canceling 
version of $X_\theta$ gates, this noise is present in those gates as well (Table~\ref{tab:gate-params}).

Finally, we model the noise afflicting the $M_Z$ and prep-$|1\>$ gates as being
much less severe than $T_2$ decoherence, motivated by a recent proposal for
an integrated charge-sensing device \cite{Gurrieri:2008a}.  Because this
device is assumed to work quickly (faster than 30\,ns) and with high sensitivity, we
model the gate time and error rate of $M_Z$ and $|1\>$ to be $\tau$ and
$p/30$ respectively.  Recall that the gate which prepares $|1\>$ is
essentially identical to the $M_Z$ gate; application of $M_Z$ returns $|0\>$
or $|1\>$, and if a $|0\>$ was mistakenly obtained, it is merely recorded
classically and managed with adaptive feed-forward correction, which appropriately re-interprets the
meaning of the state.


\section{Local Check Codes}
\label{sec:localCheckCodes}

An architecture for which transport is costly or impossible motivates the use of 
\emph{local check codes}.  Local check codes are examples of what are known as 
\emph{stabilizer codes} \cite{Nielsen:2000a}, which are characterized by a 
collection of \emph{check operators} that fix or ``stabilize'' valid code 
states.  Local check codes are also \emph{low-density parity-check} (LDPC) 
codes, namely ones in which each qubit is involved in at most a constant number 
of check operators and each check operator acts on at most a constant number of 
qubits, which eases routing requirements.  Local check codes have the additional 
property that the check operators are \emph{local} relative to some qubit 
geometry, obviating the need for transport to measure them. So far only three 
general classes of local check codes are known: surface codes 
\cite{Kitaev:2003a}, color codes \cite{Bombin:2006b}, and Bacon-Shor (BS) codes \cite{Bacon:2006a}.
In this paper we focus on BS codes because they require the 
simplest error correction circuits of these three classes. Of these, we focus on 
the simplest BS code---the one which encodes one qubit into nine (arranged in a 
$3 \times 3$ grid) that can correct for an arbitrary quantum error on a single 
qubit.  This code, which we call BS9, is depicted in Fig.~\ref{fig:BS9_21}.
For reasons explained in Sec.~\ref{sec:logicalQubit}, larger codes on a square
lattice are difficult to realize in our architecture because of transport and routing constraints.
 
BS9 quantum error correction is a two-phase process:  \emph{syndrome
extraction} followed by \emph{error recovery}.  In BS9 syndrome extraction,
one first measures the parity of every pair of horizontally-neighboring
qubits---are they of even parity ($|00\>$ or $|11\>$) or odd parity ($|01\>$
or $|10\>$)?  Then one measures the parity of the \emph{phases} of every
pair of vertically-neighboring qubits---are they of even parity ($|++\>$ or
$|--\>$) or odd parity ($|+-\>$ or $|-+\>$)?  Even if the neighboring qubits
weren't in a state of definite (bit or phase) parity before such a
measurement, they are guaranteed to be so afterwards because the measurement
of one of these ``local parity checks'' forces them to decide.  The
collection of outcomes of these measurements is called the \emph{syndrome}
because it ``diagnoses'' what, if any, errors have afflicted the encoded
state.

There are 6 vertical parity checks and 6
horizontal parity checks that need to be made.  For reasons having to do
with operator theory in quantum mechanics, we call these $XX$ and $ZZ$
checks respectively.  In principle, a single qubit can be re-used to store each of
these measurements, with the result copied off to ``classical'' storage
elsewhere between each re-use.  A more local strategy is to place
``ancilla'' or ``syndrome'' qubits in between each pair of ``data'' qubits
involved in a local check.  Since this results in 9 + 12 = 21 qubits, we
call this architecture the BS9 (21) architecture.  The syndrome and data
qubits are identified in Fig.~\ref{fig:BS9_21}.

The error recovery phase of BS9 error correction is a classical algorithm
that processes the syndrome and makes a determination of what the most
likely error is.  Because errors can afflict not just the data but also the
process of syndrome extraction itself, it is important to not be too
reactionary to the observed syndrome.  To be confident in the syndrome
values, we could repeat the process three times and take the majority of the
outcomes.  However, it turns out to be sufficient (and better) to simply
repeat once and if the two syndromes disagree, wait until the next cycle
of error correction to catch the error.  Because this method is resilient
to faults in the syndrome extraction process itself, we call it
\emph{fault-tolerant}.

Once an error is identified by the error recovery phase, it need not be
corrected immediately.  It suffices to simply keep a log of the
error and only apply the net correction when one wishes to extract
information from the logical qubit.  This removes the possibility that
errors could accumulate during the correction steps that would be applied
otherwise.  We call this the \emph{feed-forward} property of quantum error
correction.


\section{The Logical Qubit}
\label{sec:logicalQubit}
\begin{figure}[t]
	\begin{minipage}[b]{2.8in}
		\centering
		\includegraphics[trim = 0in .9in 0in 0in, clip, scale=0.34]{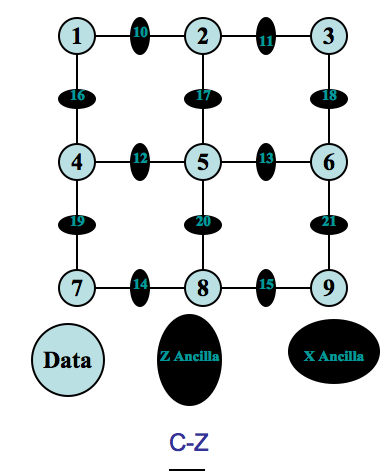}	
		\caption{Abstract architecture for the Bacon-Shor BS9 (21) code.  Horizontal
$ZZ$ parity checks are stored in $Z$-ancilla and vertical $XX$ phase parity
checks are stored in $X$-ancilla.  The lines indicate which data qubits are
involved in which parity checks.}		
		\label{fig:BS9_21}
	\end{minipage}
	\hspace{.1in}	
	\begin{minipage}[b]{3.6in}
		\centering
		\includegraphics[scale=0.25]{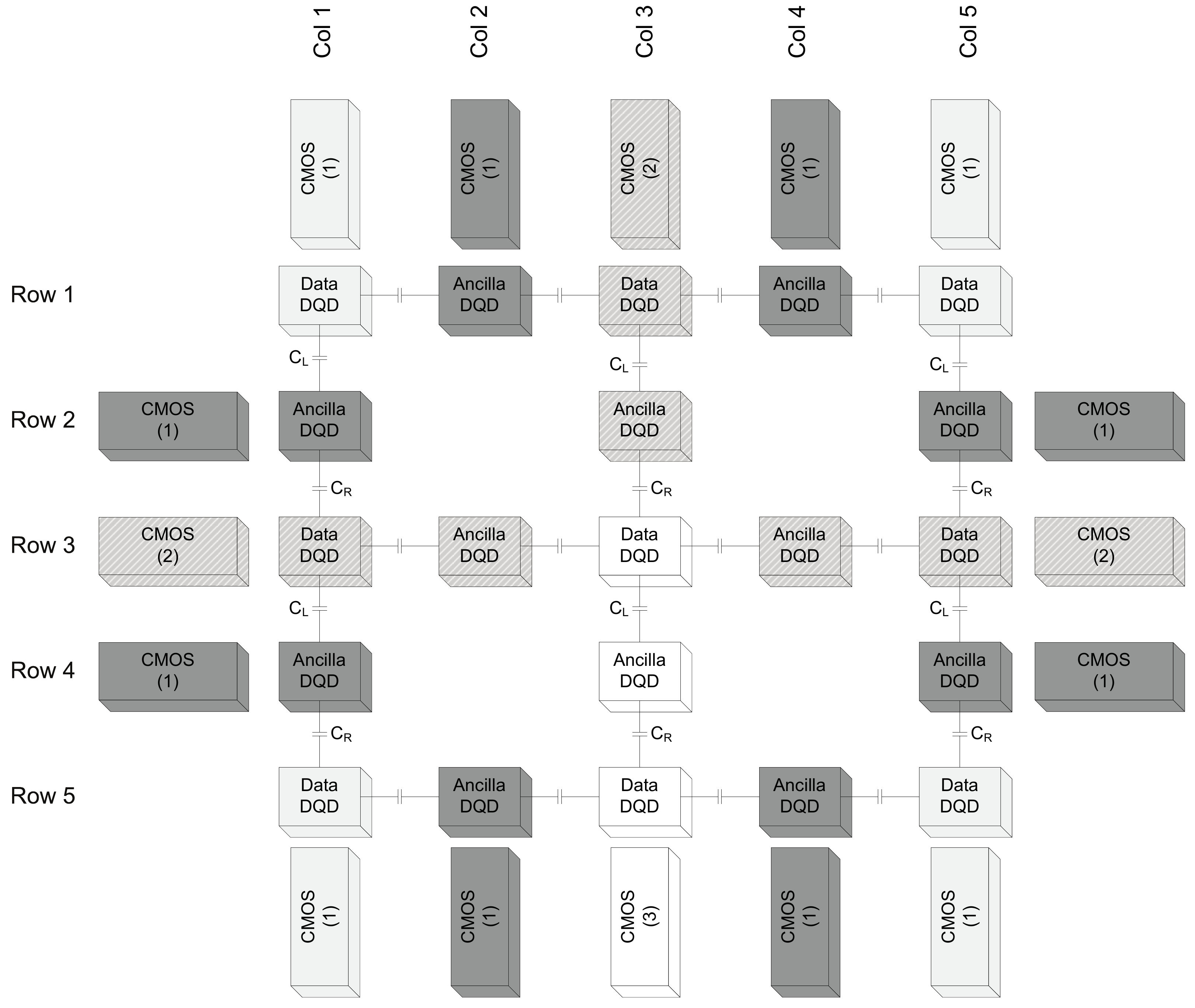}
		\caption{Enclosed 21 Qubit Bacon-Shor Architecture. Three types of CMOS control blocks, those that control: 1 DQD (light \& dark gray), 2 DQD (Gray w/ stripes) and 3 DQD (white).}
		\label{fig:BSArch}
	\end{minipage}
\end{figure}

Based on the BS9 error correction protocol, the \emph{native gate set}, and physical qubit
described in Section~\ref{sec:physicalQubit}, an ``enclosed" architecture, shown in
Fig. \ref{fig:BSArch}, was created.  The fully enclosed architecture was chosen to
optimize the parallel access of the read-out and classical electronics
to the interior qubits.  This architecture provides a platform
to probe the interactions between quantum hardware, quantum protocols, and classical
electronics for which we attempt to minimize the impact of limitations imposed 
by the classical hardware.
The architecture in Fig. \ref{fig:BSArch} consists of quantum hardware (DQDs), 
and classical CMOS electronics.  Each CMOS block controls signals to the DQDs 
shaded the same color in a particular row or column, the number of DQDs 
controlled by the CMOS block is shown in parentheses. The capacitor located 
between neighbor DQDs illustrates the electrical coupling needed to perform 2-qubit
gate operations such as \cphase.  Coupling can be done on the left or 
right side of the DQD, so vertical coupling is labeled $C_{R}$ or $C_{L}$ if the 
coupling occurs on the right or left side of the DQDs, respectively.

\begin{figure}[h]
	\begin{minipage}[b]{0.35\linewidth}
		\centering
		\includegraphics[scale=0.45]{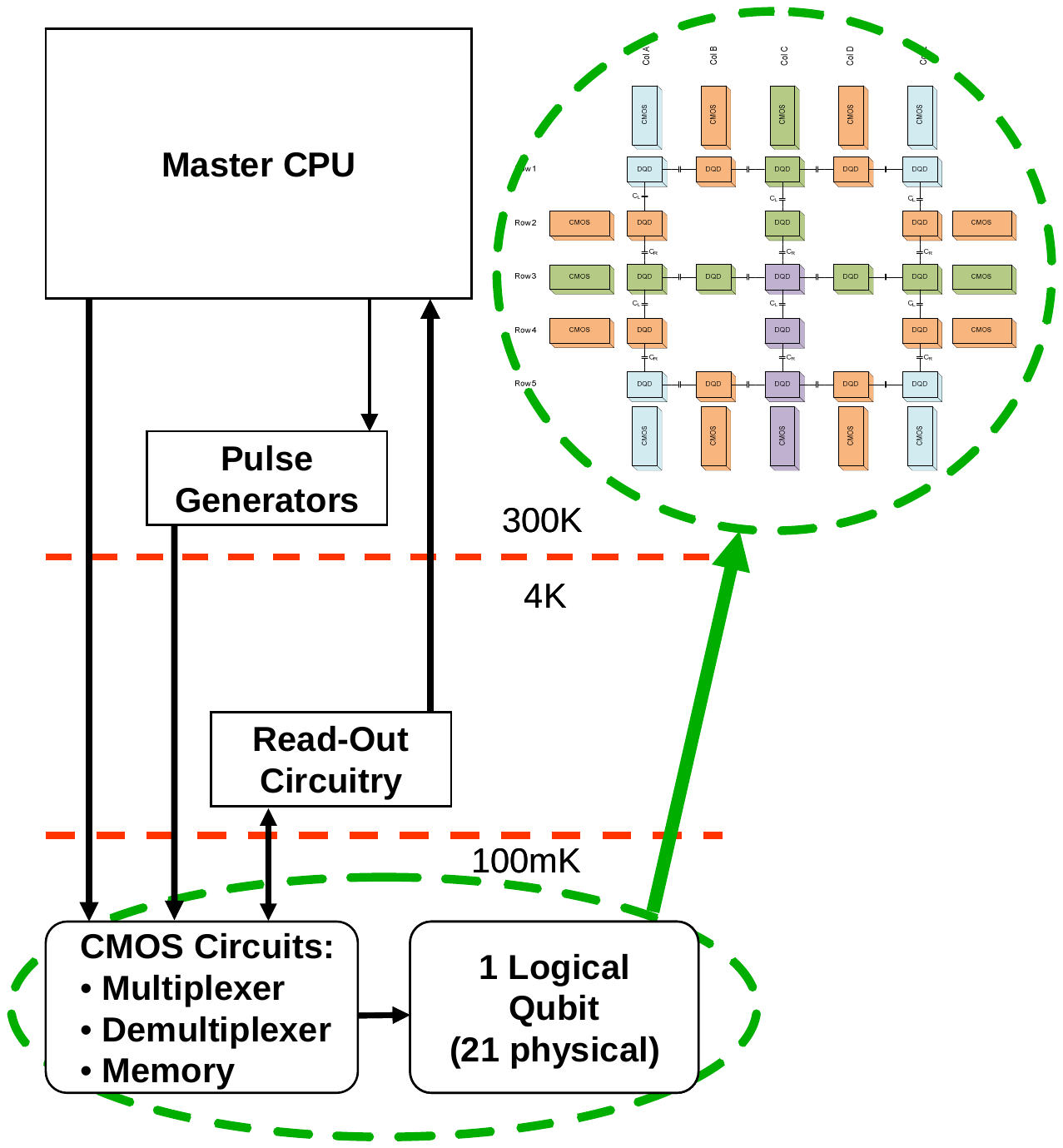}
		\caption{Cryostat Staging.}
		\label{fig:CryoStaging}
	\end{minipage}
	\hspace{0.01cm}
	\begin{minipage}[b]{0.64\linewidth}
		\centering
		\includegraphics[scale=0.4]{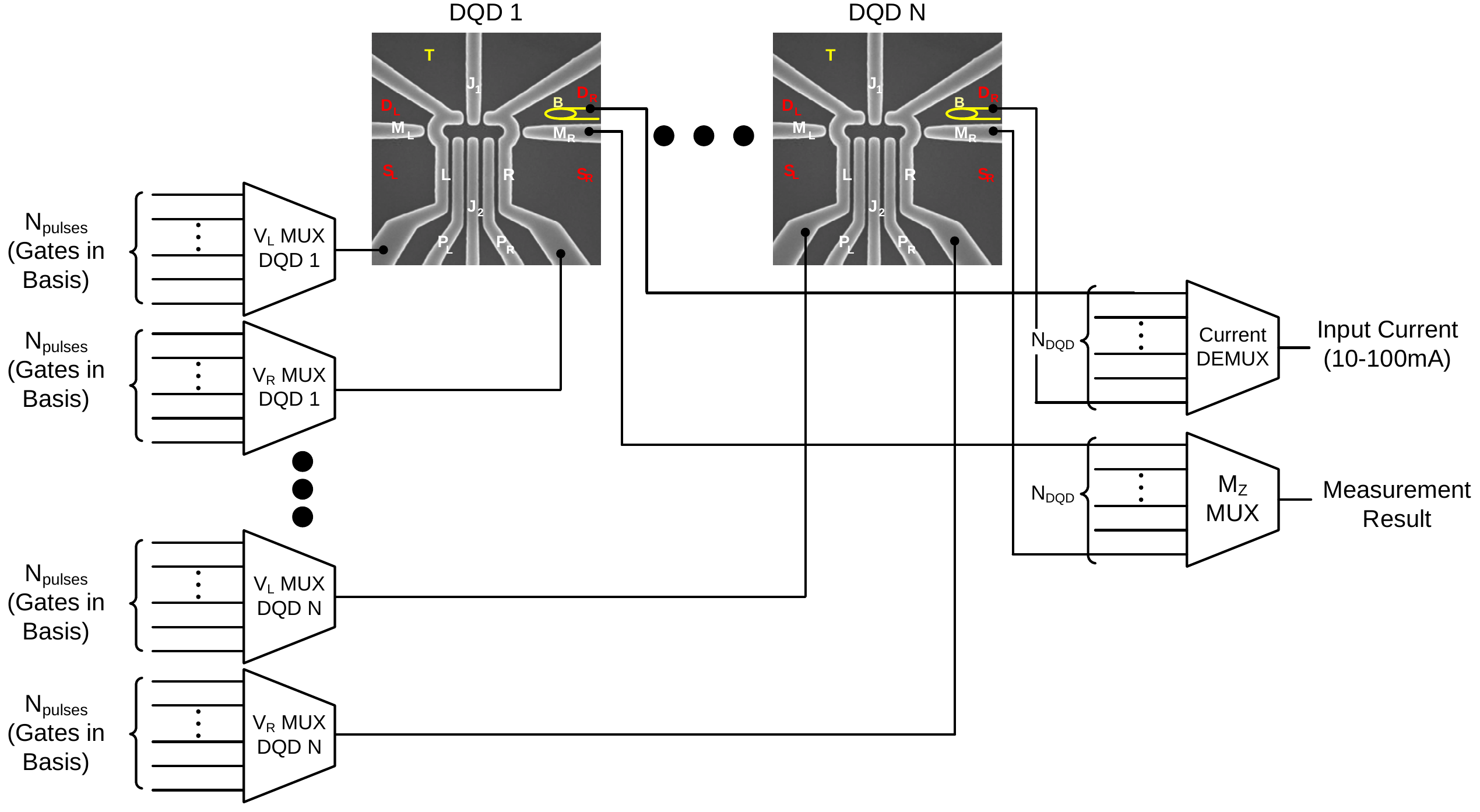}
		\caption{MUX/DEMUX within a CMOS control block (memory not shown).}
		\label{fig:DQDmuxing}
	\end{minipage}
\end{figure}

As described in Section \ref{sec:physicalQubit}, the choice of where each piece
of the classical architecture is staged in a cryostat is an important question
of trade-offs between speed, area, power dissipation, and local/global heating.
Figure \ref{fig:CryoStaging} shows the electronics staging design used in this
analysis.  The 300\,K stage holds the master CPU responsible for classical and
quantum protocol control as well as the pulse generators used to generate the
pulse sequences to gate the DQDs located at 100\,mK.  The pulse generators could
be moved to lower temperature stages, dependent on their power dissipation, but for simplicity were not in this exercise.  The
4\,K stage holds the circuitry used to readout the state of the DQDs.  The
read-out circuitry consists of a single electron charge sensor, located at 100\,mK, connected to an 
integrated comparator and latch, located at 4\,K.  The read-out is
placed at a cooling stage close to the 100\,mK stage to minimize RC delay.  The
100\,mK stage holds the DQDs and supporting classical CMOS electronics.  The CMOS
blocks at 100\,mK contain multiplexers (MUX) for routing the pulse signals from 300\,K to
each DQD, and demultiplexers (DEMUX) for reading-out the state of the DQDs (Fig.~\ref{fig:DQDmuxing}).
Additional circuitry not shown includes the memory used to hold the state of the MUX/DEMUX during
operation.

There is a limit on the number of parallel signal lines that can be run from
the higher temperature stages to the 100\,mK stage, which limits bandwidth and
parallelization.  The limitation is a result of control lines between stages
consuming area and also introducing additional heating paths between
temperatures stages.  This concern makes it desirable to use the smallest
number of connections as possible while maintaining a minimal bandwidth
penalty.  One way to achieve this is to send the control information serially
between the 300\,K stage and the 100\,mK stage.

We can analyze the number of serial lines needed to control the MUX/DEMUXs for a given number of
DQDs based on the following assumptions (Fig.~\ref{fig:DQDmuxing}):
(1) Control lines are used to set-up the MUXs and DEMUXs at
100\,mK; (2) 8 control bits are needed per DQD for this architecture; 3 bits to
setup the Right Control MUX, 3 Bits to setup the Left Control MUX, 1 Bit for Measurement
MUX, and 1 Bit for controlling the inductor current. Note the required number of control bits is a function of the number of 
gates in our \emph{native gate set}. (3) A time step is defined
as the fastest qubit gate and takes multiple serial clocks; (4) Control bits are 
pipelined.  Information being sent serially during time step ``$N$" is used during 
time step ``$N+1$" and all information must be sent before the next time step. 

\begin{figure}[h]
\centering
\includegraphics[scale=0.4]{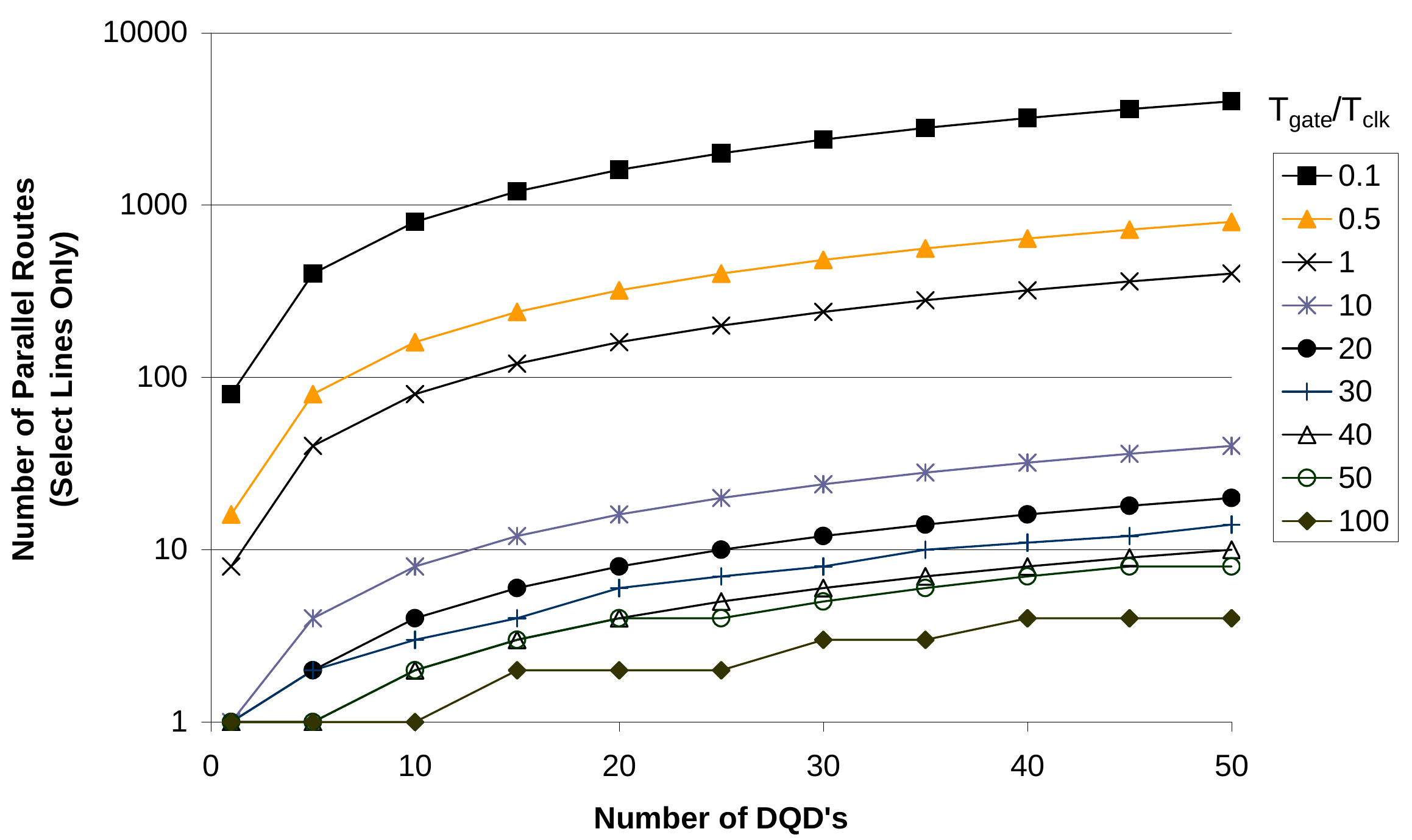}
\caption{Stage Interconnects per DQD vs. Qubit Gate to Clock Ratio.}
\label{fig:SigVsPar}
\end{figure}

The number of control lines as a function of qubits in the quantum circuit is 
calculated for several different gate to serial clock ratios, (Fig. 
\ref{fig:SigVsPar}).  A ratio of less than one implies the gate time is faster than 
the clock period, this does not apply to our architecture but is included for 
completeness. Typical cryostat's have less than 100 interconnects between 100\,mK 
and 4\,K. Several trends are highlighted by this calculation including a rapid 
and unsustainable number of lines necessary to achieve low ratios of 
T\(_{gate}\)/T\(_{clock}\).

The wire length between qubits is assumed to be as small as possible to minimize
parasitic capacitive links to other routing lines.  Unintentional
voltage variations due to cross-talk through parasitic capacitance's would lead to
unintentional qubit rotations, this is minimized with short coupling wires
between the qubits for \cphase.  For these reasons this analysis assumes that
the physical qubits must be placed very close to one another and the distance for this analysis is 250\,nm. 

The classical electronics on the 100\,mK stage that services the qubits (i.e., 
MUX-DEMUX and memory) requires non-negligible space and must be commensurate to 
the physical qubit spacing, the width of the metal lines, and the available levels of 
metal routing. The number of metal levels is limited by currently available technologies.
This imposes a constraint on the number of routing channels 
available along a single linear span of DQD qubits, Fig. \ref{fig:BSArch} 
(i.e., shaded grouping).  Because a minimum wire length between qubits is 
assumed, the CMOS blocks have a fixed width available to extend metal routes to 
service each of the DQDs.  By fixing the width of the CMOS block we 
are also fixing the number of channels available to route into the enclosed 
architecture along a single span which affects the total number of DQDs 
accessible by a CMOS block.  The total number of DQDs that can be reached for 
different technology nodes (i.e., metal widths) based on the number of shared 
or common signals among DQD is calculated in Table \ref{table:Accessibility} 
(10\,nm and 1\,nm nodes are fictional).  For this calculation the architecture 
assumes that: (1) signals are only brought from one side of the DQD, (2) the 
DQD qubit is $1\,\mu\text{m}^2$, (3) DQDs are placed 250\,nm apart, (4) each DQD 
has 17 control lines.
%
\begin{table}[h]
  \begin{center}
    \caption{Routing and Accessibility}
    \begin{tabular}{|l|c|c|c|c|c|}
    \hline    
    \bf{Technology}                         & 130\,nm & 90\,nm & 65\,nm & 10\,nm & 1\,nm\\
    \hline
    \hline                                       
    \bf{Routes per $\mu$m}                    &  19      &  27     &  40    &  462 & 4662\\
    \hline                                                     
    \bf{Accessible DQD (No Common Signal)} &  1       &   1     &   2    &  30 &  308\\
    \hline                                                     
    \bf{Accessible DQD (5 Common Signals)}  &  1       &   2     &   3    &  42 &  436\\
    \hline
    \end{tabular}
    \label{table:Accessibility}
  \end{center}
\end{table}

The number of accessible qubits is very limited even when considering relatively 
advanced CMOS nodes such as 65\,nm. This estimate, furthermore, does not 
account for a reduction of available paths due to cross talk concerns related to 
running signals directly over DQDs.  It is unclear whether these paths can be 
tolerated.  Techniques exist to reduce cross-talk such as additional ground 
planes, but this also has not been considered in this analysis, and will likely 
further decrease the number of DQDs accessible by a CMOS block.  The routing 
limitations highlight the importance of space saving approaches such as sharing 
signal lines between DQDs as well as developing more flexible lay-outs that 
provide larger spacing between DQDs or larger DQD blocks.  For the enclosed 
architecture we assume 5 common signals (Table \ref{table:Accessibility}), which 
makes it conceivable to reach 3 DQDs with one CMOS block while still using a 
non-fictional technology node (e.g., 65\,nm).  
The impact of spacing flexibility relative to increasing the number of accessible qubits provides a strong 
motivation for development of physical qubit transport methods that do not
require a high density of metal routes.

Constraints on the quantum circuit scheduling begin to emerge from these
considerations of space and 300\,K to 100\,mK bandwidth.  One goal of this work was
to establish the impact of the electronics architecture on the quantum circuit
performance.  One form in which the electronics impact manifests itself on the
performance is through establishing constraints on the QEC code schedule.  Using
the enclosed architecture as described above combined with the assumptions in
Table \ref{table:Assumptions}, a list of constraints was created and is presented in Table
\ref{table:Constraints}.
%
\begin{table}[h]
  \begin{center}
    \caption{Enclosed Architecture Assumptions}
    \label{table:Assumptions}
    \begin{tabular}{|c|p{.9\linewidth}|}
    \hline
            & \bf{Assumption} \\
    \hline                         
    \hline              
    \bf{1}  & Fastest gate will be $\sim$30\,ns for a $Z_{\pi/2}$ rotation \\
    \hline                                                     
    \bf{2}  & All DQDs are similar and tuning is not required.\\
    \hline                                                     
    \bf{3}  & 5 leads can be shared among DQDs ($T, P_L, P_R, J_1 and J_2$).\\    
    \hline
    \bf{4}  & Pulses can be applied in parallel to multiple DQDs.\\
    \hline
    \bf{5}  & Serial clock is operating at 1\,GHz\\
    \hline
    \bf{6}  & Gate pulses have the required precision to meet the error rates in \ref{tab:gate-params}. \\
    \hline
    \bf{7}  & 1 Pulse Generator exists for each gate of Table \ref{tab:gate-params}. \cphase requires 3 pulse generators.\\
    \hline           
    \end{tabular}
  \end{center}
\end{table}

\begin{table}[h]
  \begin{center}
    \caption{Enclosed Architecture Constraints}
    \label{table:Constraints}
    \begin{tabular}{|c|m{6in}|}
    \hline
            & \bf{Constraint} \\
    \hline                         
    \hline              
    \bf{1}  & Active control signals can only be passed over $Idle$ DQDs (\cphase between neighbors is acceptable).  This reduces effects of cross-talk. \\
    \hline                                                     
    \bf{2}  & The same single qubit gate can be applied to any number of DQDs controlled by a CMOS block.  Single qubit gates of differing types are not allowed.\\
    \hline                                                     
    \bf{3}  & One measurement can be done per CMOS block\\
    \hline
    \bf{4}  & \cphase gate can be applied to DQDs not controlled by the same CMOS block.\\
    \hline
    \bf{5}  & CMOS blocks controlling multiple DQDs can perform \cphase gates in parallel if the direction of coupling between DQDs is appropriate.
    				  Example: \cphase(Col3Row3, Col3Row4) \& \cphase(Col2Row5, Col3Row5) is acceptable but \cphase(Col3Row3, Col3Row4) \& \cphase(Col3Row5, Col4Row5) is not.\\
    \hline   
    \end{tabular}   
  \end{center}
\end{table}


\section{Computing an Optimal Error Correction Schedule}
\label{sec:IP}

In this section, we describe the integer program (IP) used to compute the optimal
schedule for a quantum memory using the BS9 architecture described in Section~\ref{sec:logicalQubit} and shown
in Figs.~\ref{fig:BS9_21}, \ref{fig:BSArch}, and \ref{fig:CryoStaging}.
Due to space limits, we summarize the nature of the formulation and computational considerations but do
not give a formal mathematical formulation.

We assume time is divided into ``ticks'' that represent the minimum time step 
for the shortest functional gates (\textit{I}, $I^*$, $Z_{\pi/2}$, prep 
$|1\>$, $M_Z$). All longer gates' running times are rounded to integral 
multiples of this tick size.  Thus each operation requires a small number of 
ticks.  We use a time-indexed formulation.  This allows binary decision 
variables and generally provides a tighter linear-programming relaxation, both 
important for practical solver performance.  However, the size of the 
formulation (number of variables and constraints) depends upon an upper bound 
for the schedule length, or {\em makespan}.

Our goal is to determine a legal start tick for each operation (row
and/or column circuit operation) for each qubit.
For data qubits, row/column operations cannot interleave except that $Z$-type operations at the boundary
between row and column operations
($Z_{\pi/2}$ \& \cphase) can commute.  Thus, we must always obey a series
of precedence constraints within each circuit, and we must obey
precedence constraints conditionally based on circuit ordering
decisions for each data qubit.  Qubits may be idle at any given tick.
All qubits that share a controller must perform the operation
the controller is signaling, and execute without interruption. 

We wish to schedule all the circuits subject to the constraints (Table~\ref{table:Constraints})
to minimize the total idle time of all qubits.  Holding qubits in memory contributes to qubit errors.
The data qubits are in continuous (re)use,
so their idle time is determined by the makespan.  Specifically, for
data qubit $d$, the idle time is the makespan minus the
total time to execute the gates for $d$.  The gate execution time
depends upon the number of \cphase gates, which is determined by the
qubit's physical position within its row and column circuits.
Idle time for ancilla is the total number of idle ticks between preparation and measurement.

Because operation placement and controller decision variables are
indexed by a tick, we must know a valid upper bound on the makespan
of an optimal schedule.  This determines the number of variables.
Specifically the number of variables grows linearly with the schedule
length.  Perhaps more important than the size of the formulation, the
flexibility (number of schedules) increases dramatically with
makespan, and thus solver time also increases rapidly.  It is
generally worthwhile to find a good upper bound, for example from a
valid hand-generated or heuristically-generated schedule, rather than
using a naive bound such as the length of a serial schedule.

We began by solving a version of the problem that minimizes makespan,
finding the minimum number of ticks to legally complete the schedule, rather
than the minimum total idle time.  We set the initial makespan guess from the
length of a hand-generated schedule.  By maximizing the number of
ticks where controllers are completely idle at the end of the schedule,
we can compute a minimum makespan.  We then used the minimum makespan
as the first estimate for the makespan when computing a minimum-idle-time schedule.  
This objective explicitly trades off makespan vs. ancilla idle time.  The optimal
schedule for this makespan had only two total idle ticks summed over all
ancilla.  Because increasing the makespan would add $9$ ticks of idle time
taken over all the data bits, while possibly only remove $2$ ticks from
the ancilla, this schedule is optimal over all makespans.

For practical performance, we computed tight legal time windows on all
the operation variables and a reasonable lower bound on the makespan.
We ran the IP using the ampl mathematical programming language and the
cplex 11.0 integer programming solver on a dual-core 32-bit linux
workstation with 3.06Ghz Xeon processors with 2Gb of RAM. The final
problem, to compute the minimum idle time knowing the optimal makespan
solved in less than a second.  The problem ampl sent to the solver
after preprocessing had 4701 binary variables, 5784 constraints, and
58163 nonzeros. The IP computed an optimal schedule that had 95 idle ticks, 
which is a significant improvement from a hand-generated schedule with 
129 idle ticks.

Adding DD requires significant changes.  We enumerate the
possible DD blocks for each qubit, built from a set of maximal
DD blocks.  We require the IP to select precisely one
DD block containing any operation that requires DD.
For data qubits, each tick must be inside a DD block.  Idle
time at the end of the ``wait'' interval of the last block can wrap
around to cover idle time at the start of the schedule. The controller
must signal the $X$-gates used for DD as appropriate given the block
starting time.  We constrain the operations to run during the
appropriate ``wait'' interval for their DD block.
We allow the previous constraints to enforce precedence and all other operation constraints.
Computing optimal DD-based schedules for \textit{Z}-$Idle$-\textit{Z}-$Idle$ refocusing is future work.


\section{BS9 Threshold Calculations}
\label{sec:threshold}
In order to assess the performance of any logical qubit architecture, we need an 
explicit noise model.  Here we consider the \emph{depolarized noise} (DPN) model 
in which each gate other than measurement is chosen to work flawlessly with 
probability $1-p$ or else be followed by one of the non-identity Pauli operators 
selected uniformly at random.  Sometimes we will consider a variant of this 
noise model, the \emph{biased} DPN model, in which certain gates are more or 
less likely to fail than others.  We generally draw the relative error 
probabilities for the biased DPN model from Table \ref{tab:gate-params} in 
Sec.~\ref{sec:physicalQubit} to model classical electronics constraints, 
although sometimes we tweak this biasing to study the relative importance of 
certain gates' fault rates.  While we estimate $p = 0.3\%$ for realistic gates 
in Table~\ref{tab:gate-params}, we consider more general DPN models in this section in which $p$ is 
variable. We note that in both noise models, in addition to measurements, the 
identity gate (representing idling qubits) will have a fault probability different 
from the rest of the gates.  Furthermore the failure probability of the identity 
gate will be a constant and will depend on whether or not DD is used during 
error correction.  This is done to maintain consistency with our hardware 
characteristics described in Section 
\ref{sec:physicalQubit}. 

The accuracy threshold is a figure of merit for a logical qubit architecture 
realizing arbitrary computation \cite{Aliferis:2006a}.  In this work we are not 
interested in arbitrary computation, rather just the creation of a logical 
qubit memory.  Consequently we define a figure of merit called the error threshold that 
is closely related to the accuracy threshold, but is more relevant to our work. 
Assume the noise in the error correction circuit is DPN with failure probability 
$p$, then the probability of failure for error correction [$P_{EC}(p)$] is an 
increasing function of $p$.    
\newline
 {\bf Approximate Definition}:    \emph{The error threshold for a DPN model is
  the maximum failure probability $p_{th}$ for which $p \le p_{th} \Rightarrow P_{EC}(p) \le p$.}
\newline
The accuracy threshold is defined in a similar way with the maximum failure 
probability of the logical gate set replacing ($P_{EC}()$) in the above 
definition.  In our work there is only one logical gate, namely the logical 
identity with failure probability $P_{EC}(p)$.  However, there is a problem with 
the above definition in our context.  The identity gate in our error 
correction circuit has a constant failure probability ($p_I>0$) and as such 
$P_{EC}(0)>0$.  Hence, there is no $p_{th}$ that satisfies the above definition. 
We want to highlight this nature of $P_{EC}(p)$, since it is significantly 
different from the fault tolerant logical gate failure probabilities for 
distance-3 codes \cite{Aliferis:2006a} that behave like $O(p^2)$.  We reiterate that 
this difference is due to the fact that the idle periods in our architecture 
have a fixed non-zero error probability.  With this in mind and allowing for 
a biased DPN we define:
\newline
{\bf Definition}:  \emph{The error threshold for a (biased) DPN model is the 
maximum failure probability $p_{th}$ for which there exists an $\epsilon >0$ 
such that $p \in [p_{th}-\epsilon,  p_{th}] \Rightarrow P_{EC}(p) \le cp$, where 
$cp$ is the maximum probability of failure of the native gate (NG) set.}  Note that 
$c=1$ for DPN and $c=4$ for the biased DPN from Table \ref{tab:gate-params}.
\newline
The intuition behind the above definition is the following.  Error 
correction is realized using the native gate set while accounting for
hardware constraints.  The error threshold is the failure probability that the native 
gates need to perform better than in order for the error correction failure probability to 
be lower (better) than the worst individual gate used in the error correction. 
We note that in the context of designing for a logical qubit, one could also define an 
error threshold based on memory.  In other words, assuming a noise model like DPN, 
one could ask for the input error probability $p$ such that the \emph{failure rate} of 
error correction is lower (better) than the \emph{failure rate} of an idle.  Here 
\emph{failure rate} implies the failure probability is normalized by the time 
taken for the "gate".  In view of space limitations we do not present our 
work in this context.

A threshold only exists when syndrome extraction is processed
fault-tolerantly, such as via the ``repeat twice'' syndrome measurement
strategy we suggested in Sec.~\ref{sec:localCheckCodes}, or more generic
strategies such as the ones proposed by Shor \cite{Shor:1996a}, Steane
\cite{Steane:1997a} and Knill \cite{Knill:2004a}.  Regardless of which
fault-tolerant syndrome extraction strategy one uses, there is a
well-developed, but technical method using ``extended rectangles'' (EX-REC) \cite{Aliferis:2006a} that
allows one to estimate the threshold of any architecture.  

Using the EX-REC method, we computed the error threshold using Monte
Carlo techniques for the BS9 (21) code for the following settings, each one becoming increasingly closer
to describing an architecture constrained by the limitations of classical
electronics:

\begin{enumerate}

\item Syndrome measurements are not modeled as circuits but rather as
``black box'' processes that either occur ideally or suffer a fault
according to a DPN model.

\item Syndrome measurements are effected by circuits subject to a DPN model
that implement Steane's fault-tolerant protocol \cite{Steane:1997a}, where
these circuits are expressed over the (nonlocal) gate basis $\{|0\>, |1\>,
\CNOT, M_X, M_Z\}$.

\item Syndrome measurements are effected by IP-optimized circuits described
in Sec.~\ref{sec:IP}, constrained by the classical electronics requirements
listed in Sec.~\ref{sec:logicalQubit}, implementing the ``measure twice''
fault-tolerant protocol described in Sec.~\ref{sec:localCheckCodes} over the
Si DQD native gate basis described in Sec.~\ref{sec:physicalQubit}.
Computing the error threshold in this setting is one of the main goals of
this paper.  Both the biased and unbiased DPN models are considered.  
The entire analysis is done for both the situations in which ($a$) there is no DD and with a provably 
optimum schedule for minimizing idles ($b$) there is $X-Idle-X-Idle$ DD \emph{without} an optimal schedule.  The 
change in error correction using DD from the no DD case is that there are 
additional $X$ and $I$ gates, but the failure probability of the identity drops 
from $1 \times 10^{-2}$ to $5 \times 10^{-7}$.

\end{enumerate}

To perform Monte Carlo studies of the error threshold $p_{th}$ for these
settings, we used a self-modified extension of the QDNS simulator
\cite{Imre:2004a} to input an EX-REC for the error correction circuit of the desired form
for the BS9 (21) code.  We then fed this circuit into a
message-passing-interface parallelized Monte Carlo simulator we developed
that estimates the failure probability of the error correction circuit [$P_{EC}(p)$] in
the DPN and biased DPN models.  Finally, we computed the error threshold using a
bisection search to determine the crossover point, i.e., when 
$P_{EC}(p_{th})=p_{th}$ for DPN and $P_{EC}(p_{th})=4p_{th}$ for the biased DPN 
models.  The results of our Monte Carlo studies are presented in 
Table~\ref{tab:MC-thresholds}.
\begin{table}[h]
\centering
\begin{tabular}{|c|c|c|c|c|} \hline
 & Black box & Steane &NG \& Opt. Sched. & NG, DD, Sub-Opt. Sched. \\ \hline \hline
$p_{M_Z} = 0$ & $1.7 \pm 0.1 \%$ & & & \\ \hline
$p_{M_Z} = 2p/3$ & $1.1 \pm 0.1 \%$ & & & \\ \hline
$p_{I} = 10^{-2}$ & & $7.3 \pm 0.1 \times 10^{-4}$ & & \\ \hline
$p_{I} = 10^{-5}$ & & $1.1 \pm 0.1 \times 10^{-3}$ & & \\ \hline
$p_{I} = 0$ & & $1.1 \pm 0.1 \times 10^{-3}$ & & \\ \hline
unbiased & & &  No Threshold  & $5.5 \pm 0.1 \times 10^{-5}$ \\ \hline
biased & & &  No Threshold  & $2.0 \pm 0.1 \times 10^{-5}$ \\ \hline
\end{tabular}
\caption{Monte-Carlo accuracy threshold estimates for various syndrome
measurement models.  Each model assumes that all of the gates are equally
likely to fail unless otherwise specified in the left column.}
\label{tab:MC-thresholds}
\end{table}

Several trends can be extracted from this table.  To begin, the threshold is
very high in the ``black-box'' model, reflecting the fact that this model
just probes the quantum error-correction code's ability to correct
data errors independent of the architecture surrounding it.  The
$p_{M_Z} = 0$ case examines precisely this; the $p_{M_Z} = 2p/3$ case is
motivated by the fact that only $X$ and $Y$ errors cause an $M_Z$
measurement to fail where a $Z$ error does not.  The Steane model results show 
that gate errors in the syndrome extraction can contribute much more to the 
threshold than data errors during idles, causing a drop in the threshold by a 
factor of more than 10 even when idles are error free ($p_I = 0$).  In part, 
this is an artifact of the nonlocal gates in the Steane circuit which can be 
parallelized to such a degree that they hardly leave data qubits idle at all. The NG 
model shows the large penalty incurred in changing the circuitry to accommodate 
classical electronics constraints faced by a Si DQD architecture.  Even with an 
optimal schedule to minimize the number of idles we find there is no threshold.  This is 
due to the large number of idles in the NG model (Table \ref{tab:BS9gates}) 
whose error rates are significantly high ($1 \times 10^{-2}$), resulting in 
$P_{EC}(0) \approx 0.4$.  In other words, even if the native gate set is error 
free the error correction has a failure probability of 0.4. However, when we 
incorporate DD into the error correction we vastly improve the situation and 
actually obtain an error threshold.  This is because although DD increases the 
number of $X$ and $I$ gates in the error correction, which would lead one to 
assume increases $P_{EC}(p)$, it more than compensates for this by decreasing 
$p_I$ by a factor of $T_2/T_2^* = 60\,\text{ms}/3\,\mu\text{s} = 2000$.  This 
suggests that under the current hardware assumptions for our architecture, DD is 
essential.  Finally, while the error threshold for biased DPN is lower than the 
error threshold for unbiased DPN, it is not significantly smaller and in fact 
may even be more achievable.  This is because the biased DPN allows some gates 
to be less reliable than others; for example, the $2.0 \times 10^{-5} $ error 
threshold for the biased DPN requires that $X_{\pi/2}$, $X$, and $\CPHASE$ 
gates err with probability below $8.0 \times 10^{-5}$, which is a higher 
probability than than the $5.0 \times 10^{-5}$ error threshold demanded for 
these gates by the unbiased DPN.       

\begin{table}[h]
\centering
\begin{tabular}{|c|c|c|}
\hline
Gate                		 & BS9(21) w/o DD & BS9(21) with DD  \\ \hline
 \hspace{1 cm}     	 & $\#$ of Gates   & $\#$ of Gates    \\ \hline \hline
Prep $|1\rangle$ & 12  	        & 12  	\\ \hline
$X_{\pi/2} $		 & 42 		        & 42 		\\ \hline
$Z_{\pi/2} $		 & 18		        &18		  \\ \hline
$X$         		 & 0			        & 104   \\ \hline
$\cphase$        & 24		        & 24    \\ \hline
$M_Z$       		 & 12 	          & 12    \\ \hline
$I$ / $I^*$      & 95			      &  219  \\ \hline
\end{tabular}
\caption{Gate count for realizing error correction using the native gate set in our architecture}
\label{tab:BS9gates}
\end{table}


\section{Conclusion}
\label{sec:conclusion}
This paper describes a solid-state error-corrected logical qubit memory that
accounts for constraints imposed by both electronics and the native gate
set. The combination of physical qubit geometry, electronics layout, and
lack of a viable transport mechanism forces a restricted qubit layout and
positioning of supporting electronics. We chose the Bacon-Shor code, a local
check code, to accommodate the transport constraint. We also observe that
limits on space for metal routing, even for the 65~nm CMOS process nodes,
result in a maximum reach of three DQDs per CMOS block.  This
drove us to consider the smallest Bacon-Shor code that can correct for a
single-qubit error, the BS9 code.

We developed a fault-tolerant procedure for carrying out error-correction
with the BS9 code for our memory architecture, including a schedule for the
native gate set that was constrained by limits of the electronics.  We
constructed rules for limits on simultaneous parallel gate operations and
gate time to reflect concerns regarding cross-talk and signal bandwidth
limits between cryostat stages.  The electronics scheduling constraints lead
to extra idle time penalties.  We created a general IP
methodology for minimizing idles in error-correction subject to the hardware
constraints.

We developed a figure of merit relevant for our architecture that we call the
\emph{error threshold} that closely resembles the accuracy threshold for
fault-tolerant quantum memory in the literature.  Using this figure of merit
and a 30 ns clock period, which leads to very error-prone idles, we find
that our architecture has no error threshold, even with a provably-optimal
schedule. 
 
The idle error probability can be suppressed dramatically with dynamical
decoupling (DD) pulses.  Using $X$-\textit{Idle}-$X$-\textit{Idle} DD 
and a hand-generated non-optimal error-correction schedule, we
found that even with a 30 ns clock period, our architecture achieved an error
threshold of $5.5 \times 10^{-5}$ for a depolarized noise model and $2.0
\times 10^{-5}$ for a biased noise model.  This highlights the importance of
combining multiple error-suppression strategies when classical electronics
constraints are considered.  We also find that these threshold values are a
factor of approximately twenty larger than more abstract architectures.  This is
principally because of the extra 104 $X$ gates required for DD during
error correction.  Additional idles are also added to the schedule due to
DD, but DD significantly reduces the impact of these idles.

In summary, we examined a hypothetical solid-state logical qubit memory
architecture constrained by more realistic electronics and native gate set
constraints.  Layout constraints motivate a local error correction code
choice. Electronics constraints on scheduling manifest themselves as a
penalty due to additional idle times, which in turn requires additional
gating and DD to suppress idle noise.  We calculated optimal schedules and
an error threshold for this more realistic case. We note that although this
analysis was specific to a silicon solid-state implementation and quantum memory, the insight
and tools developed apply more generally, especially to other implementations
that are intended to operate in cryostats with CMOS electronics control and
read-out. 

{\footnotesize{


}}

\end{document}